\documentclass{aa}
\newcommand{\rin}{r_1}
\newcommand{\rout}{r_2}
\newcommand{\qB}{\vec{q_\mathrm{B}}}
\newcommand{\tchi}{\tilde{\chi}}
\newcommand{\vvv}{\vec{\varv}}

\newcommand{\vphi}{\varphi}

\newcommand{\bvfreq}{N_0}
\newcommand{\omegaT}{M_0}
\newcommand{\dd}{\mathrm{d}}

\newcommand{\id}{\mathrm{i}}

\newcommand{\ddr}{\frac{\dd}{\dd r}}

\newcommand{\hatb}{\hat{\vec{b}}}
\newcommand{\vnabla}{\vec{\nabla}}

\newcommand{\er}{\hat{\vec{e}}_{r}}
\newcommand{\etheta}{\hat{\vec{e}}_{\theta}}
\newcommand{\ephi}{\hat{\vec{e}}_{\vphi}}
\newcommand{\nel}{n_{\mathrm{e},1}}
\newcommand{\kB}{k_\mathrm{B}}

\newcommand{\rtp}{r_\mathrm{c}}
\newcommand{\cV}{{\mathcal{V}}}

\newcommand*\volave[1]{\langle#1\rangle_{\cV}}
\newcommand{\rvir}{R_\mathrm{vir}}
\newcommand{\vvir}{\varv_\mathrm{vir}}

\newcommand{\omegaTin}{M_1}
\newcommand{\omegaTout}{M_2}

\newcommand{\U}{r^2\rho_0\delta\varv_r}

\makeatother
\DeclareDocumentCommand{\sect}{ m O{} O{} O{} O{} O{} O{} }{%
  \IfEmptyTF{#1}%
    {Sect.}%
    {\IfEmptyTF{#2}%
      {Sect. \ref{sec:#1}}%
      {Sect. \ref{sec:#1}%
        \IfEmptyTF{#3}%
        { \andet\ \ref{sec:#2}}%
        {, \ref{sec:#2}%
         \IfEmptyTF{#4}%
           {, \andet\ \ref{sec:#3}}%
           {, \ref{sec:#3}%
           \IfEmptyTF{#5}%
             {, \andet\ \ref{sec:#4}}%
             {, \ref{sec:#4}%
             \IfEmptyTF{#6}%
               {, \andet\ \ref{sec:#5}}%
               {, \ref{sec:#5}%
               \IfEmptyTF{#7}%
                 {, \andet\ \ref{sec:#6}}%
                 {, \ref{sec:#6}, \andet\ \ref{sec:#7}}%
               }%
             }%
           }%
        }%
      }%
    }%
}
\makeatletter
\def\IfEmptyTF#1{%
  \if\relax\detokenize{#1}\relax
    \expandafter\@firstoftwo
  \else
    \expandafter\@secondoftwo
  \fi}

\usepackage{iflang}
\newcommand\andet{%
  \IfLanguageName{english}%
                 {and}%
                 {et}%
  \ignorespaces}
\newcommand\eqeq{%
  \IfLanguageName{english}%
                 {Eq}%
                 {Éq}%
  \ignorespaces}
\makeatother
\DeclareDocumentCommand{\eq}{ m O{} O{} O{} O{} O{} O{} }{%
  \IfEmptyTF{#2}%
    {\eqeq. (\ref{eq:#1})}%
    {\eqeq s. (\ref{eq:#1})%
      \IfEmptyTF{#3}%
      { \andet\ (\ref{eq:#2})}%
      {, (\ref{eq:#2})%
       \IfEmptyTF{#4}%
         {, \andet\ (\ref{eq:#3})}%
         {, (\ref{eq:#3})%
         \IfEmptyTF{#5}%
           {, \andet\ (\ref{eq:#4})}%
           {, (\ref{eq:#4})%
           \IfEmptyTF{#6}%
             {, \andet\ (\ref{eq:#5})}%
             {, (\ref{eq:#5})%
             \IfEmptyTF{#7}%
               {, \andet\ (\ref{eq:#6})}%
               {, (\ref{eq:#6}), \andet\ (\ref{eq:#7})}%
             }%
           }%
         }%
      }%
    }%
}
\makeatother
\DeclareDocumentCommand{\eqnp}{ m O{} O{} O{} O{} O{} O{} }{%
  \IfEmptyTF{#2}%
    {\eqeq. \ref{eq:#1}}%
    {\eqeq s. \ref{eq:#1}%
      \IfEmptyTF{#3}%
      { \andet\ \ref{eq:#2}}%
      {, \ref{eq:#2}%
       \IfEmptyTF{#4}%
         {, \andet\ \ref{eq:#3}}%
         {, \ref{eq:#3}%
         \IfEmptyTF{#5}%
           {, \andet\ \ref{eq:#4}}%
           {, \ref{eq:#4}%
           \IfEmptyTF{#6}%
             {, \andet\ \ref{eq:#5}}%
             {, \ref{eq:#5}%
             \IfEmptyTF{#7}%
               {, \andet\ \ref{eq:#6}}%
               {, \ref{eq:#6}, \andet\ \ref{eq:#7}}%
             }%
           }%
         }%
      }%
    }%
}

\usepackage{natbib}
\bibpunct{(}{)}{;}{a}{}{,}

\defcitealias{latter15}{LFF15}
\defcitealias{pl22a}{PL22a}
\defcitealias{pl22b}{PL22b}
\defcitealias{kempf24}{KR25}
\defcitealias{kempf23}{Kempf et al. 2023}

\usepackage{xcolor}
\usepackage{multirow}
\usepackage{graphicx}
\usepackage{txfonts}
\usepackage[breaklinks, colorlinks, citecolor=blue, urlcolor = blue]{hyperref}
\begin{document} 

   \title{Global linear analysis of the magneto-thermal instability in a stratified spherical model of the intracluster medium}

   \author{J. M. Kempf
          \inst{1}
          \and
          H. Latter
          \inst{2}
          }

   \institute{$^1$Institut de Recherche en Astrophysique et Planétologie (IRAP), Université de Toulouse, CNRS, Toulouse, France\\
              \email{jean.kempf@irap.omp.eu}\\
              $^2$Department of Applied Mathematics and Theoretical Physics (DAMTP), University of Cambridge, Cambridge, United Kingdom
}

   \date{\today}

 
  \abstract
  {
  The buoyancy stability properties of dilute plasma, as found in the intracluster medium (ICM),
  are dramatically modified
  because of the anisotropic transport of heat along the magnetic field lines.
  This feature gives rise
  to the magneto-thermal instability (MTI) when the temperature gradient is aligned with the gravity,
  which systematically occurs in the outskirts of galaxy clusters.
  }
  {
  Most previous linear analyses of the MTI adopted a local approach and the Boussinesq formalism.
  However, the conduction length, which sets the characteristic length scale of the MTI,
  might be a non-negligible fraction of the scale height in the ICM.
  We want to assess the impact of locality assumptions on the linear physics of the MTI.
  Another goal is to unveil the deeper connections between
  these global MTI modes and
  their magneto-rotational instability (MRI) counterparts in accretion discs.
  Our third objective is to provide a new benchmark against which any numerical code implementing
  the Braginskii heat flux in spherical geometry can be tested.
  }
  {
  We perform a global linear analysis
  of the MTI in a spherical stratified model of the ICM, subject to a Navarro-Frenk-White gravitational potential
  of dark matter.
  We use a combination of analytical results from both the Sturm-Liouville theory
  and WKBJ approximations,
  corroborated by numerical results obtained with both a pseudo-spectral Chebyshev solver
  and the finite-volume code IDEFIX,
  to better explain the physics of the global MTI eigenmodes.
  }
  {
  We obtain scaling laws and approximate expressions for the growth rates of the global modes.
  We show that the associated eigenfunctions are confined within an inner region,
  limited by a turning point,
  where the mode is allowed to grow.
  The most unstable local MTI modes correspond to the portion of the global mode
  localised near the turning point.
  This phenomenology is very similar to that of the global MRI modes in accretion discs.
  Finally, direct numerical simulations successfully reproduce the global MTI modes and their growth rates,
  with errors smaller than 1\%.
  }
  {
  Overall, this study provides us with new insights on the linear theory of the global MTI in the ICM,
  and a useful numerical test bench for any astrophysical fluid dynamics code
  embedding anisotropic heat flux.
  }


   \titlerunning{Global linear analysis of the MTI in a stratified spherical model of the ICM}
   \maketitle


\section{Introduction}
\label{sec:intro}

  Galaxy clusters are the largest gravitationally bound objects in the universe,
  with a typical length scale being the virial radius, $\rvir\sim \mathrm{Mpc}$,
  and with a total mass ranging from $10^{14}$ to $10^{15}$ solar masses \citep{girardi98}.
  The gravitational potential well maintaining the internal coherence of such massive structures
  originates from yet-to-be-detected dark matter \citep{zwicky33,smith36},
  which is responsible for ${\sim}85\%$ of their total mass \citep[see for example][and references therein]{mcnamara07}.
  Galaxies, observable in the optical wavelengths, only represent ${\sim}3\%$ of this mass.
  The remaining ${\sim}12\%$ mass is found as a tenuous and very hot plasma,
  usually referred to as the intracluster medium (ICM), that shines in X-rays \citep{fabian03,fabian06}.
  Typical values of density and temperature in the ICM are
  $\rho\sim10^{-27}\ \mathrm{g/cm^3}$ and $\kB T\sim 5\ \mathrm{keV}$, respectively,
  where $\kB$ is the Boltzmann constant.
  Radio observations of galaxy clusters also reveal that these astrophysical objects
  are magnetised up to $B\sim 10\ \mathrm{\mu G}$ \citep{ferrari08}.

  From a plasma physics standpoint,
  such measurements allow for the computation of typical plasma parameters \citep{nrl}.
  For instance, the Larmor radius of charged particles in the ICM reaches no more than the nanoparsec scale,
  while their collisional mean free path can reach
  the kiloparsec scale \citep{kunz22}.
  On the macroscopic level, the subsequent ‘dilute’ regime
  of the ICM plasma
  induces an anisotropic transport of heat, along magnetic field lines only \citep{braginskii65},
\begin{equation}
  \qB = -\kappa \hatb\hatb\cdot\vnabla T,
\label{eq:bhf}
\end{equation}
  usually referred to as the Braginskii heat flux.
  In this equation, $\hatb=\vec{B}/B$ is the unit magnetic field vector and  $\kappa$ is the thermal conductivity,
  which should not be confused with the thermal diffusivity,
  $\chi=\kappa/\left(\rho\mathcal{R}\right)$,
  $\mathcal{R}$ being the specific gas constant.
  The anisotropic heat flux triggers a new class of buoyancy instabilities, possibly relevant to the ICM:
  the magneto-thermal instability (MTI) where the background temperature is decreasing with radius \citep{balbus00}
  and the heat-flux buoyancy driven instability (HBI) in case of an increasing background temperature \citep{quataert08}.
  As seen with X-ray observations, the ICM systematically features negative temperature gradients
  in outer cluster regions,
  from a typical cooling radius for relaxed clusters \citep{leccardi08,simionescu11,ghirardini19},
  which are thus unstable to the MTI in principle.
 \citet[][respectively, PL22a, PL22b, KR25, hereafter]{pl22a,pl22b,kempf24} showed that the non-linear saturation of the MTI can drive fluid turbulence
  at levels roughly consistent with those deduced from X-ray observations of galaxy clusters
  \citep{hitomi16,hitomi18,xrism25a,xrism25b}.
  The characterisation of such turbulence is critical for our understanding of the ICM physics:
  it could induce non-thermal pressure support \citep{pratt19}, turbulent heating \citep{zhuravleva14b}, and magnetic field amplification \citep{donnert18,rincon19},
  to name but a few of the most outstanding problems related to galaxy clusters.

  Both the MTI and the HBI are local instabilities that share a common maximum growth rate,
  $\omegaT\sim\mathrm{Gyr^{-1}}$.
  The local dispersion relation for MTI elevator modes
  (i.e. those with zero vertical wavenumber)
  was originally found by \citet{balbus00}:
\begin{equation}
  D\left(\sigma, k_x\right) = \sigma^3 + \tchi k_x^2 \sigma^2 + \bvfreq^2 \sigma - \tchi k_x^2  \omegaT^2 = 0,
\label{eq:cartdispbalbus}
\end{equation}
  where $\sigma$ and $k_x$ are respectively the growth rate and the horizontal wavenumber of the local MTI mode,
  while $\tchi$ and $\bvfreq$ are respectively the reduced thermal diffusivity and the Brunt-Väisälä frequency,
  defined later in \sect{model}.
  Equation (\ref{eq:cartdispbalbus}) is only valid for local elevator modes,
  whose typical length scales are much smaller than any ICM scale height.
  However, the MTI is known to preferentially develop at the conduction length,
  $\ell_\chi = \sqrt{\chi/\omegaT}$ \citepalias{pl22a,pl22b},
  which can be a non-negligible fraction of the cluster size itself \citepalias{kempf23,kempf24}.
  At these scales, appreciable variation of the background ICM thermodynamic profiles necessarily happens.
  Consequently, the validity of local models may be called into question
  when describing the linear stability and nonlinear saturation of the MTI.
  While \citetalias{kempf24} previously addressed the question of the non-linear saturation
  thanks to global simulations of magneto-thermal turbulence,
  a global linear analysis of the MTI remains to be carried out
  \citep[for a first approach to this problem though, see Appendix B in][]{berlok21}.
  Such a study would help to further understand why
  the phenomenology developed by \citetalias{pl22a,pl22b} for MTI-driven turbulence
  in local ICM atmosphere successfully applies to global simulations too \citepalias{kempf24}.

  In the case of the HBI, previous global linear analyses unveiled interesting vertical structures
  of the corresponding eigenmodes \citep{latter12,berlok16b};
  but they did not establish any formal link between global modes and their local counterparts.
  On the other hand, \citet[][LFF15 hereafter]{latter15} carried out a global linear analysis
  of the magneto-rotational instability (MRI) in accretions discs,
  and elucidated the relationship between local and global modes of the MRI.
  As first intuited by \citet{balbus01},
  the local stability properties of the MTI and the MRI are very much alike.
  Both instabilities modify, in the same way, the classical stability criteria of the gas.
  Therefore, we build a global linear theory of the MTI,
  drawing from that of \citetalias{latter15} for the MRI,
  exploiting their common linear features.
  The main goal of this paper is to shed light on the connection between local and global MTI modes,
  and between the global linear MRI and MTI too.

  Eventually, the non-linear study of such instabilities at saturation relies on large, complex,
  compressible, MHD codes for the simulation of astrophysical flows.
  However, only few detailed analytical benchmarks are found in the literature for global curvilinear geometries \citepalias{latter15},
  especially so for numerical operators as exotic as the anisotropic Braginskii heat flux \citep{latter12, berlok16b}.
  A global linear analysis of the MTI in the ICM would therefore provide an additional test, in spherical geometry,
  to validate the computational fluid dynamics tools on which a lot of theoretical astrophysics relies nowadays.

  The outline of the paper is as follows.
  In \sect{model}, we first describe the ICM model,
  along with the analytical and numerical methods, used to carry out a global linear analysis of the MTI.
  In \sect{analytical}, we present several new interesting analytical results
  (some of which are deferred to Appendix \ref{app:sl}).
  We derive the radial structures of global MTI eigenmodes
  and unveil their deeper relationships with local MTI modes,
  and with global MRI eigenmodes in accretion discs.
  In \sect{numerical}, these analytical results are further corroborated
  by numerical solutions obtained with both a Chebyshev pseudo-spectral solver
  and direct numerical simulations performed with the finite-volume code IDEFIX.
  Our main conclusions are summarised in \sect{conclusion}.

\section{Model and methods}
\label{sec:model}

  In this section, we first introduce the fully compressible Braginskii-MHD (B-MHD) equations.
  We then present the equilibrium model of a stratified ICM,
  around which we carried out a global linear analysis of the MTI.
  Finally, we describe the numerical methods used both to solve the subsequent eigenproblem
  and to run direct numerical simulations.

\subsection{Model of the intracluster medium}
\subsubsection{The compressible Braginskii magnetohydrodynamic equations}
  The B-MHD framework
  provides evolution equations for the density, momentum, and (internal) energy
  of a dilute, collisional plasma subject to anisotropic transport or heat and/or momentum.
  Our B-MHD equations read
\begin{align}
\label{eq:densitycons}
  &\frac{\partial \rho}{\partial t}
  + \vnabla \cdot (\rho \vvv)
  = 0, \\
\label{eq:momentumcons}
  &\frac{\partial (\rho \vvv)}{\partial t}
  + \vnabla \cdot \left(\rho \vvv\vvv
  + p \vec{I}
  + \frac{B^2}{2\mu_0} \left(\vec{I}
  - 2\hatb\hatb\right)
  \right)
  =
  \rho \vec{g}, \\
\label{eq:Eint}
  &\frac{\mathcal{R}}{\gamma-1}\left[\frac{\partial \left(\rho T\right)}{\partial t}
  + \vnabla \cdot \left(\rho T \vvv \right)\right]
  = 
  - \vnabla\cdot\qB 
  - p\vnabla\cdot\vvv, \\
\label{eq:induction}
  &\frac{\partial \vec{B}}{\partial t}
  = \vnabla \times (\vvv \times \vec{B}).
\end{align}
  All the symbols here have their usual meanings, which are made explicit in \citetalias{kempf24}.
  The gravitational acceleration, denoted $\vec{g}$, is specified in the next paragraph.
  The heat flux, $\qB$, in a dilute, collisional plasma
  is given by \eq{bhf}.
  In this analysis, we dismissed any non-ideal effects (such as isotropic viscosity or magnetic resistivity)
  except for the anisotropic transport of heat.
  Specifically, we neglected the Braginskii viscosity,
  which otherwise accounts for the effects of anisotropic pressure \citep{snyder97}.
  Besides, we ignored any plasma effects (such as kinetic micro-instabilities)
  that could affect the anisotropic heat flux on which the MTI relies \citep{berlok21, perrone24a, perrone24b}.
  The system of \eq{densitycons}[momentumcons][Eint][induction] was closed thanks to
  the equation of state for a perfect gas with adiabatic index $\gamma$.

\subsubsection{Initial ICM atmosphere at hydrostatic equilibrium}
  Since our goal was to carry out a global linear analysis of the MTI
  in the spherical equatorial plane, $\theta = \pi/2$, of a stratified ICM,
  we therefore introduced a 2D spherical coordinate system, $(\er,\ephi)$.
  The position vector is $\vec{r} = r\er + \vphi\ephi$.
  The domain extends from an inner radius, $\rin$, to an outer radius, $\rout$,
  and from $0$ to $2\pi$ in azimuth, $\vphi$.
  We took advantage of the global spherical geometry at hand to set up a background ICM atmosphere
  as realistic as possible.
  In galaxy clusters, the gravitational potential well of dark matter is well described by 
  the Navarro-Frenk-White potential \citep[NFW;][]{navarro97}.
  We used a NFW gravitational acceleration field, $\vec{g} = -g_0 \er$, of intensity
\begin{equation}
  g_0(r) = \left(\frac{\vvir^2}{\rvir}\right)
           \frac{1}{r^2}\frac{\log\left(1+cr\right) - cr/\left(1+cr\right)}{\log\left(1+c\right) - c/\left(1+c\right)},
\label{eq:nfw}
\end{equation}
  where $r$ is naturally rescaled to the virial radius, $\rvir$,
  and $c$ is a concentration parameter that we set to $c=5$.
  We also introduced the virial velocity,
  $\vvir = \left(GM_\mathrm{vir}/\rvir\right)^{1/2}$,
  where $M_\mathrm{vir}$ is the virial mass (not to be confused with the MTI frequency, $\omegaT$, in \eqnp{cartdispbalbus}).

  We considered a simple, purely azimuthal, initial magnetic field,
  $\vec{B} = B_0 \ephi$, with $B_0$ uniformly weak and independent of the radius, $r$, across the domain.
  This magnetic configuration is the most unstable to the MTI.
  Our choice was driven by pragmatism and tractability.
  In a 2D equatorial model of an ICM atmosphere, purely azimuthal fields are 
  the only possible magnetic configurations unstable to the MTI,
  also ensuring the thermal equilibrium of any spherically symmetric temperature profile,
  since the background radial heat flux is initially shut off.

\begin{figure}
\includegraphics[width=0.98\hsize]{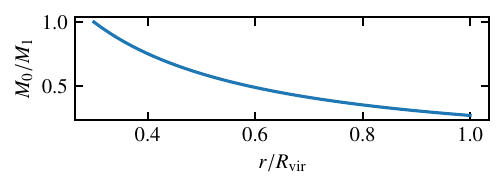}
\caption{
  Local MTI frequency profile, $\omegaT$, as a function of the radius, $r$.
}
\label{fig:profileMTIfrequency}
\end{figure}

  The temperature profile, $T_0(r)$, is then a free parameter.
  Given the theoretical nature of the work undertaken here,
  we favoured a simple analytic function, rather than a complex fit to the observational data
  \citep{ghirardini19,kempf23}.
  We adopted a temperature profile
  at thermal equilibrium in the presence of a hypothetical background heat flux
  (with isotropic conductivity, or equivalently, anisotropic conductivity and purely radial magnetic field),
  a reasonable compromise between physical plausibility and analytical tractability.
  A thermal equilibrium in the presence of isotropic heat flux and uniform thermal conductivity is obtained by setting
\begin{equation}
  T_0(r) = T_1 + \left(T_2-T_1\right) \frac{\rin - r}{\rin - \rout} \frac{\rout}{r},
\end{equation}
  subject to boundary conditions (BCs) $T(\rin)=T_1$ and $T(\rout)=T_2 < T_1$,
  at the inner and outer radii, $\rin$ and $\rout$, respectively.
  We set the values of these free parameters,
  based on observations of the Perseus cluster \citep{simionescu11}, to
  $\kB T_1 = 6\ \mathrm{keV}$ and $\kB T_2 = 2\ \mathrm{keV}$,
  for $\rin = 0.3\rvir$ and $\rout = \rvir$, respectively.
  For Perseus, we have $\rvir=1.1\ \mathrm{Mpc}$.
  Other virial quantities are given in \citetalias{kempf24}.
  As shown in Fig. \ref{fig:profileMTIfrequency},
  with such choices, the resulting profile of MTI frequency is a well-behaved, monotonically decreasing
  function of the radial coordinate,
\begin{equation}
  \omegaT(r) = \sqrt{-g_0\frac{\dd \log T_0}{\dd r}}.
\end{equation}
  The MTI frequency at the innermost radius, $\rin$, is
  $\omegaTin = 2.44\ \mathrm{Gyr^{-1}}$.
  Since the profile is decreasing with radius, it is also the maximum MTI frequency within the domain.
  For further reference, we also define $\omegaTout$, the (smallest) MTI frequency
  at the outer boundary, $\rout$.

  Finally, a thermal diffusivity profile, $\chi(r)$,
  needs to be specified too.
  Strictly speaking, this physical quantity should be
  the Spitzer diffusivity, $\chi_\mathrm{S}\propto T^{5/2}/\rho$, in the ICM plasma \citep{spitzer62}.
  However, we show in \sect{analytical} that
  the precise dependency of the thermal diffusivity only plays a modest role,
  at least in the limit of fast thermal diffusion that we introduce later.
  Therefore, so as not to complicate the interpretation
  of both the analytical derivations and numerical results presented in the next sections,
  we used a uniform profile of thermal diffusivity,
  set to the Spitzer value it would have at the inner boundary,
\begin{equation}
\label{eq:tchi}
  \chi_{\mathrm{s},1} = 0.11 \rvir\vvir
  \left(\frac{\kB T_1}{6 \ \mathrm{keV}}\right)^{\frac{5}{2}} \left(\frac{\nel}{10^{-3}\ \mathrm{cm^{-3}}}\right)^{-1},
\end{equation}
  where $\nel$ is the electronic density number, close to $10^{-3}\ \mathrm{cm^{-3}}$ at $\rin=0.3\rvir$
  in the Perseus cluster \citep{simionescu11}.
  As a result, the density profile, $\rho_0(r)$, needs not be explicitly computed
  to solve the perturbation equations, introduced in the next subsection.

\subsection{Global linear analysis}
  In order to carry out a global linear analysis of the MTI in the ICM,
  the compressible B-MHD \eq{densitycons}[momentumcons][Eint][induction]
  were linearised to first order around the previous background HSE.
  All perturbed quantities were azimuthally expanded into monochromatic Fourier modes
  since this direction is naturally periodic in spherical geometry
  and the background state is spherically symmetric.
  As a starting point for this study, we discarded any dependency of global MTI modes on the polar angle, $\theta$,
  and only considered radial and azimuthal variations.
  Their radial structure was then captured into a function of the radius only:
  $\delta f(t, r,\vphi) = \delta f(r)\exp\left(\sigma t +\id m \vphi\right)$.
  This ansatz is appropriate to describe a 2D equatorial cut of MTI modes
  with large scale variation in the polar direction, but smaller scale (though still global) variations
  in the other directions.
  Moreover, guided by the local theory, we knew that the main features of the MTI derive
  in a 2D framework, so that the $\theta$-variation can be safely neglected.
  This also leaves the problem analytically tractable.
  Accordingly, the amplitude of our initial magnetic field, $B_0$, is independent
  of the polar coordinate, $\theta$, in the 2D equatorial cut.
  As discussed in \sect{conclusion},
  the cost for moving beyond this approximation would be to solve a 2D, rather than a 1D, eigenproblem
  because the Braginskii heat flux is a non-linear operator.
  A very similar approximation was recently used by \citet{choudhury25},
  to linearly study cold fronts in a 2D equatorial ICM plane.
  
\subsubsection{Perturbation equations}
  From the linearisation at first order of \eq{densitycons}[momentumcons][Eint][induction],
  the perturbation equations are
\begin{align}
\label{eq:drho}
\sigma \frac{\delta\rho}{\rho_0}&= -\frac{\id m}{r}\delta \varv_\vphi - \frac{1}{r^2\rho_0}\frac{\dd}{\dd r} \left(r^2\rho_0 \delta \varv_r\right),  \\
\label{eq:dvphi}
\sigma\delta \varv_\vphi&=  -\mathcal{R} T_0 \frac{\id m}{r} \frac{\delta p}{p_0}, \\
\label{eq:dvr}
\sigma\delta \varv_r&=  - \mathcal{R}T_0\frac{\dd}{\dd r}\left(\frac{\delta p}{p_0}\right) + g_0 \frac{\delta T}{T_0}, \\
\label{eq:dT}
\sigma \frac{\delta T}{T_0} &= - \left(\gamma-1\right) \left[\frac{\id m}{r} \delta\varv_\vphi + \frac{1}{r^2}\ddr\left(r^2 \delta\varv_r\right)\right] - \frac{\dd\log T_0}{\dd r} \delta\varv_r \\
&- \gamma\tchi\frac{m^2}{r^2}\left(\frac{\delta T}{T_0}-\frac{\dd \log T_0}{\dd r} \frac{\delta A}{B_0}\right), \nonumber \\
\label{eq:dA}
\sigma \frac{\delta A}{B_0} & = - \delta \varv_r,
\end{align}
  where we introduced the reduced thermal diffusivity, $\tchi(r)=\chi(r)\left(\gamma-1\right)/\gamma$.
  The pressure perturbation is simply deduced from the perturbed equation of state,
\begin{equation}
  \frac{\delta p}{p_0} = \frac{\delta\rho}{\rho_0} + \frac{\delta T}{T_0}.
\label{eq:idealgasane}
\end{equation}
  The perturbed potential vector, $\vec{\delta A}=\delta A \etheta$, relates
  to the perturbed magnetic field through $\vec{\delta B} = \vnabla\times\vec{\delta A}$.
  In the linearised \eq{dvphi}[dvr] for the velocity perturbations,
  we ignored the feedback from the magnetic field on the flow through the action of the Lorentz force.
  This simplification is justified when the plasma beta parameter is very large, that is in the limit $\beta\gg 1$.
  This assumption holds to some extent for the plasma of the ICM.
  In this context, the only dynamical role of the magnetic field is to channel heat through
  the Braginskii heat flux, \eq{bhf}.
  The set of linearised equations above is solved numerically in \sect{numerical},
  but is slightly simplified, in the next subsection, for the need of the analytical derivations
  carried out in \sect{analytical}.

\subsubsection{Sound-proof approximation}

  Before the global linearised \eq{drho}[dvphi][dvr][dT][dA] can be reworked into
  a simple and useful ordinary differential equation (ODE),
  we needed to simplify \eq{drho}.
  We adopted a type of anelastic model, in which we assumed that the flow is very subsonic
  and that the fractional thermodynamic perturbations remain small.
  Doing so allowed us to drop the time derivative of the density perturbation, $\sigma \delta\rho/\rho_0$, in \eq{drho},
\begin{equation}
  0 = \frac{\id m}{r}\delta \varv_\vphi + \frac{1}{r^2\rho_0}\frac{\dd}{\dd r} \left(r^2\rho_0 \delta \varv_r\right).
\label{eq:drhomod}
\end{equation}
  This approximation filters out sound waves that would otherwise interfere with the computation
  of the linear MTI modes, in which we were interested.
  This approach is justified as long as the Mach number of the flow is small enough\footnote{Equivalently, the condition (that the flow is very subsonic, $\mathcal{M}\ll1$) for the anelastic approximation can be rewritten in terms of frequency as $\sigma\ll\omega_\mathrm{s}\sim c_\mathrm{s}/H$, with the sound frequency, $\omega_\mathrm{s}$, and the length scale at which the mode develops, $H$, because $\sigma\sim\delta\varv/H\sim\omega_\mathrm{s}\mathcal{M}$.},
  that is $\mathcal{M} = \delta\varv/c_\mathrm{s}\ll 1$, with the sound speed, $c_\mathrm{s} = \sqrt{\gamma\mathcal{R}T_0}$.
  This is guaranteed for arbitrarily small velocity perturbations.
  We stress that this procedure is not strictly equivalent to an anelastic approximation,
  at least not as usually found in solar physics \citep{miesch05},
  even though they look very similar.
  The main difference here is that we do not require the background temperature gradient
  to be the adiabatic gradient \citep{rieutord15},
  to within a small parameter.
  Indeed, the ICM is stable to thermal convection, precisely because this background gradient
  is usually much lower than the adiabatic gradient in galaxy clusters \citep{cavagnolo09}.

\subsection{Numerical methods}
  We first describe the Chebyshev pseudo-spectral method designed to solve the full eigenproblem
  \eq{drho}[dvphi][dvr][dT][dA].
  We then present the Godunov-type code IDEFIX for astrophysical flows,
  used in \sect{numerical} to cross-check the radial structure and growth rate
  of the analytically derived global MTI eigenmodes.

\subsubsection{Chebyshev pseudo-spectral solver}
  We discretised \eq{drho}[dvphi][dvr][dT][dA] using a pseudo-spectral Chebyshev method \citep{boyd01} on a Gauss-Chebyshev grid.
  Only two BCs need to be specified because, as shown in \sect{analytical},
  these equations can be rewritten as a second-order ODE.
  Dirichlet BCs were then imposed on the radial velocity field,
  through basis recombination of the Chebyshev polynomials.
  The eigenvalues, $\sigma_n$, of the resulting matrix were computed with SCIPY \citep{virtanen20}.
  The eigenvalue problem was then solved at fixed azimuthal order, $m$,
  for different $n$-indexed modes which are decreasingly ranked according to their growth rates.
  Our solver was adapted from previous quasi-global linear analyses of either the HBI or the MTI \citep{latter12,berlok21},
  and was successfully tested against those.
  We did not use a single grid resolution for all computations.
  Instead, we required that the variation of the numerically computed eigenvalue of the eigenmode
  under consideration did not exceed a tolerance threshold of $10^{-5}$ when the grid resolution was doubled.
  The number of grid points needed for the numerical convergence
  of most modes presented here usually did not exceed $512$.
  However, it was increased up to $2048$ when we needed to solve for very large mode numbers, up to $n\approx1000$, in \sect{numericalcheby}.

\subsubsection{Direct numerical simulations with IDEFIX}
  IDEFIX is a new Godunov-type finite-volume code for astrophysical fluid dynamics \citep{lesur23}.
  It incorporates a Braginskii module, which includes in particular the anisotropic heat flux, \eq{bhf}.
  In this work, we used IDEFIX to solve the fully compressible B-MHD \eq{densitycons}[momentumcons][Eint][induction].
  We note however that IDEFIX solves the conservation equation for the total energy \citepalias[][\eqeq. 9]{kempf24}
  instead of the conservation equation for the internal energy, \eq{Eint}.
  The numerical methods used in all the simulations presented in \sect{numerical}
  are the following: piece-wise linear reconstruction with a Van Leer slope limiter,
  HLLD Riemann solver, three-stage Runge-Kutta time integrator,
  super time-stepping Runge-Kutta-Legendre scheme \citep{meyer14,vaidya17}
  for the Braginskii heat flux \citep[no slope limiter though,][]{sharma07a},
  and upwind constrained transport contact algorithm \citep{londrillo04}.
  Even though the simulations were ran in spherical geometry,
  we only solved the equations in a 2D domain here, that is
  the equatorial plane, $(\er,\ephi)$, at $\theta=\pi/2$,
  extending from $\rin = 0.3$ to $\rout = 0.7$ in radius and from $0$ to $2\pi$ (or from $0$ to $\pi$)
  in the azimuthal direction.
  When a smaller azimuthal extent is used, from $\vphi=0$ to $\pi$, periodic BCs remain applicable
  provided that the azimuthal order, $m$, of the initial condition is even and greater than or equal to $2$
  because $e^{\id m \vphi}$ is $2\pi/m$-periodic, and, thus, at least $\pi$-periodic.
  Accordingly, the domain extent used in IDEFIX was always smaller in radii
  (and sometimes in azimuth too) than that used in the pseudo-spectral solver.
  The reason behind this choice was to ensure that IDEFIX solved the global MTI eigenmodes
  with enough radial (respectively azimuthal) resolution,
  when the azimuthal order, $m$, (resp. mode number, $n$), of the modes became large.
  The numerical resolution used was therefore $2048\times4096$.
  The initial parameters were $\chi=0.11$ (in line with \eqnp{tchi})
  and $B_0=10^{-6}$, that is $\beta\sim10^{12}$.

  To obtain a true static equilibrium upon which the MTI can grow, along the lines of \eq{drho}[dvphi][dvr][dT][dA],
  we chose an ICM atmosphere initially at HSE.
  However, imposing a HSE at machine precision in any Godunov-type codes, such as IDEFIX,
  is difficult \citep{zingale02},
  because HSE is numerically obtained via the (only approximate) cancellation of two possibly large numbers:
  the pressure gradient and the gravitational pull.
  If not done properly, or without sufficient radial resolution,
  a spherically symmetric, $m=0$, acoustic wave can develop.
  If the amplitude of this acoustic perturbation exceeds the initial amplitude
  of the seeded MTI eigenmodes, the latter can be disrupted,
  and unwanted modes with other azimuthal orders might appear.
  To avoid such pathological behaviours,
  we made use of a balance scheme, available in IDEFIX.
  The goal of this module is to enforce the initial condition,
  which needs to be an analytic equilibrium, as a true numerical equilibrium.
  It does so by initially computing, and then removing at each time step,
  the right-hand side numerical residual of the initial condition,
  so that the analytical equilibrium is recovered at machine precision.
  All simulations performed in \sect{numerical} employed this balance scheme.
  The discussion about the BCs used in these simulations is deferred there,
  since they depend on the specific radial structure of the MTI eigenmodes, which we obtain later. 

\section{Analytical results}
\label{sec:analytical}
  In this section, we present our analytical results 
  on the global linear MTI eigenproblem.
  First, we elucidated the single ODE governing the MTI eigenmodes,
  and we applied it a specific limit of fast thermal diffusion to obtain
  a problem that is to some extent analytically tractable.
  Then, we qualitatively discuss the physical phenomenology of the global MTI modes
  induced by the simplified Schrödinger potential of the governing ODE.
  More quantitative results were obtained too, thanks to known theorems from the Sturm-Liouville theory;
  but they are deferred to Appendix \ref{app:sl} to avoid breaking the flow of the current section.
  Finally, we obtained convenient analytical approximations to the growth rates of the MTI eigenmodes,
  using WKBJ approximations in the limits of large azimuthal orders, $m/n \gg 1$, or large mode numbers, $n/m \gg 1$.
  The detailed derivations leading to these approximate growth rates
  are also deferred to Appendix \ref{app:crude}.

\subsection{Single governing ordinary differential equation}
\label{sec:ode}

  First, we transformed \eq{dT},
  using \eq{drho}[idealgasane], so as to introduce the Brunt-Väisälä frequency,
  defined as
\begin{equation}
  \bvfreq^2(r) = \frac{g_0}{\gamma}\frac{\dd \log\left(p_0\rho_0^{-\gamma}\right)}{\dd r},
\end{equation}
  in:
\begin{align}
\sigma \left(\frac{\delta T}{T_0} - \frac{\gamma-1}{\gamma}\frac{\delta p}{p_0}\right) &= -\frac{\bvfreq^2}{g_0}\delta \varv_r \\  &- \tchi \frac{m^2}{r^2}\left(\frac{\delta T}{T_0}-\frac{\dd \log T_0}{\dd r} \frac{\delta A}{B_0}\right).
\label{eq:dTmod}
\end{align}
  The latter equation is exact, with respect to \eq{dT}. No further approximation was used to derive it.
  The sound-proof \eq{drhomod},
  combined with \eq{dvphi}[dvr][dA][dTmod],
  can then be reworked into a single second-order ODE, for the variable $r^2\rho_0\delta\varv_r$,
\begin{align}
  \label{eq:schro}
  \left[
  \frac{\dd^2}{\dd r^2}\right.
  &+ \frac{1}{H}\frac{\sigma/\gamma + \tchi m^2/r^2}{\sigma + \tchi m^2/r^2}\ddr \\
  &-\left. \frac{m^2/r^2}{\sigma^2\left(\sigma + \tchi m^2/r^2\right)} D\left(\sigma,m,r\right)
  \right]
  \left(r^2\rho_0\delta v_r\right) = 0, \nonumber
\end{align}
  where we introduced the pressure scale height,
\begin{equation}
  H = - \left(\frac{\dd \log p_0}{\dd r} \right)^{-1} > 0.
\end{equation}
  The operator $D$ is defined similarly to the local elevator MTI dispersion relation, \eq{cartdispbalbus},
  though adapted to the spherical geometry at hand,
\begin{equation}
  D\left(\sigma, m/r\right) = \sigma^3 + \tchi(r) \frac{m^2}{r^2} \sigma^2 + \bvfreq^2(r) \sigma - \tchi(r) \frac{m^2}{r^2} \omegaT^2(r).
\label{eq:sphdispbalbus}
\end{equation}
  If $m/r$ is formally replaced by a horizontal wavenumber $k_x$ within the equation $D\left(\sigma, m/r\right) = 0$,
  we recover the original dispersion relation from \citet{balbus00}, \eq{cartdispbalbus}, for local elevator modes
  (i.e. $\dd/\dd r = 0$).
  Moreover, if the derivative operator, $\dd/\dd r$, is locally approximated by a radial wavenumber, $k_r$,
  and in the limit $k_r H \gg 1$ (which may be valid for modes with shortly varying radial structure),
  we then regain from \eq{schro} the dispersion relation for arbitrary local 2D perturbations
  \citep[][\eqeq. 18, for vanishing viscosity]{parrish12a}.

\subsection{Fast thermal diffusion limit and Sturm-Liouville problem}
  Unfortunately, the eigenproblem, \eq{schro},
  cannot be recast as a Sturm-Liouville ODE,
  which prevented us from applying results from this theory.
  Further theoretical progress was then only achievable at the cost of an additional hypothesis.
  Fast thermal diffusion (i.e. $\tchi m^2/r^2 \gg \sigma, \omegaT, \bvfreq$) is a common limit often used in studies of
  magnetised buoyancy instabilities, such as the MTI and the HBI \citep{balbus00,quataert08}.
  Under this assumption, \eq{schro} can formally be reworked into
\begin{align}
\label{eq:ouioui}
  \left[
  \frac{\dd^2}{\dd r^2}
+ \frac{1}{H}\ddr
- \frac{m^2}{r^2} 
  \right]
  \left(r^2\rho_0\delta\varv_r\right) =
- \frac{m^2}{r^2} \frac{\omegaT^2(r)}{\sigma^2}\left(r^2\rho_0\delta\varv_r\right).
\end{align}
  For our ICM model however, we have $\ell_\chi \sim H \lesssim \rvir$, therefore $\tchi m^2/r^2 \gg \omegaT$
  also means $\ell_\chi^{-2} \sim H^{-2} \ll m^2/r^2$.
  In other words, the limit of fast thermal diffusion corresponds to modes with length scales much smaller
  than the pressure scale height, $H$,
\begin{equation}
  \left|\frac{\dd^2}{\dd r^2}\left(r^2\rho_0\delta\varv_r\right)\right| \sim \left|\frac{m^2}{r^2}\left(r^2\rho_0\delta\varv_r\right)\right| \gg \left|\frac{1}{H}\frac{\dd}{\dd r}\left(r^2\rho_0\delta\varv_r\right)\right|.
\end{equation}
  This fact is later used to approximate the growth rate, $\sigma$, with a WKBJ procedure,
  with $m^{-1}$ as the small control parameter.
  The first-order derivative term can then be dropped in \eq{ouioui},
  which can directly be rewritten as a Sturm-Liouville equation,
\begin{align}
  \left[
  \frac{\dd^2}{\dd r^2}
- \frac{m^2}{r^2} 
  \right]
  \left(r^2\rho_0\delta\varv_r\right) =
- \epsilon \frac{m^2}{r^2} \frac{\omegaT^2(r)}{\omegaTin^2}\left(r^2\rho_0\delta\varv_r\right),
  \label{eq:ftdsl}
\end{align}
  with homogeneous Dirichlet BCs, $\delta\varv_r(\rin)=\delta\varv_r(\rout)=0$.
  The weight function is
  $-(m^2/r^2)(\omegaT^2(r)/\omegaTin^2)$,
  with $\omegaTin$ the MTI frequency at the innermost radius,
  and the eigenvalue is $\epsilon = \omegaTin^2/\sigma^2$.
  The structure of this problem ensures a discrete set of real eigenvalues, $\epsilon$,
  that we order increasingly: $\epsilon_{n} < \epsilon_{n+1}$ for $n \in \mathbb{N}$.
  With this convention, the smallest Sturm-Liouville eigenvalue, $\epsilon_0$, is reached for
  the highest growth rate, $\sigma_0$.
  Besides, the $n^\mathrm{th}$ eigenfunction associated to the eigenvalue $\epsilon_n$
  possesses $n$ zeroes in the corresponding solution domain (excluding those at the boundaries).

  In Appendix \ref{app:sl}, we make use of the explicit Sturm-Liouville form of \eq{ftdsl},
  in the limit of fast thermal diffusion, along with known results from the theory,
  to further constrain both the growth rate and the radial structure of the global MTI eigenmodes.
  Here, we simply state the obtained results, which are twofold.
  First, the growth rate, $\sigma$, of any global eigenmode is bounded by the maximum MTI frequency, $\omegaTin$,
  in the domain, which is the growth rate of the fastest local elevator mode.
  Second, the global MTI eigenmodes are exactly the functions that maximise
  their power weighted by the maximum growth rate of local MTI modes, quantified by
\begin{equation}
  R\left[u\right] = \int_{\rin}^{\rout} \frac{m^2}{r^2}\frac{\omegaT^2(r)}{\omegaTin^2} \left|u\right|^2\dd r,
\label{eq:R}
\end{equation}
  under the constraint that the quadratic form,
\begin{equation}
  Q\left[u\right] = \int_{\rin}^{\rout} \left(\frac{m^2}{r^2} \left|u\right|^2 + \left|\frac{\dd u}{\dd r}\right|^2\right)\dd r,
\label{eq:Q}
\end{equation}
  is kept constant.
  The latter \eq{Q} induces a natural subsequent norm, $Q$, for the MTI eigenfunctions,
  which is used to normalise any global MTI modes presented in what follows, so that $Q\left[\U\right]=1$.
  This mathematical optimality result is well in line with the physical phenomenology of the MTI eigenfunctions,
 dictated by the presence of a turning point at a critical finite radius, which we describe next.

\subsection{Physics of the global MTI eigenmodes}
\label{sec:tp}

\subsubsection{Turning points -- global and local eigenmodes}

\begin{figure}
\includegraphics[width=\hsize]{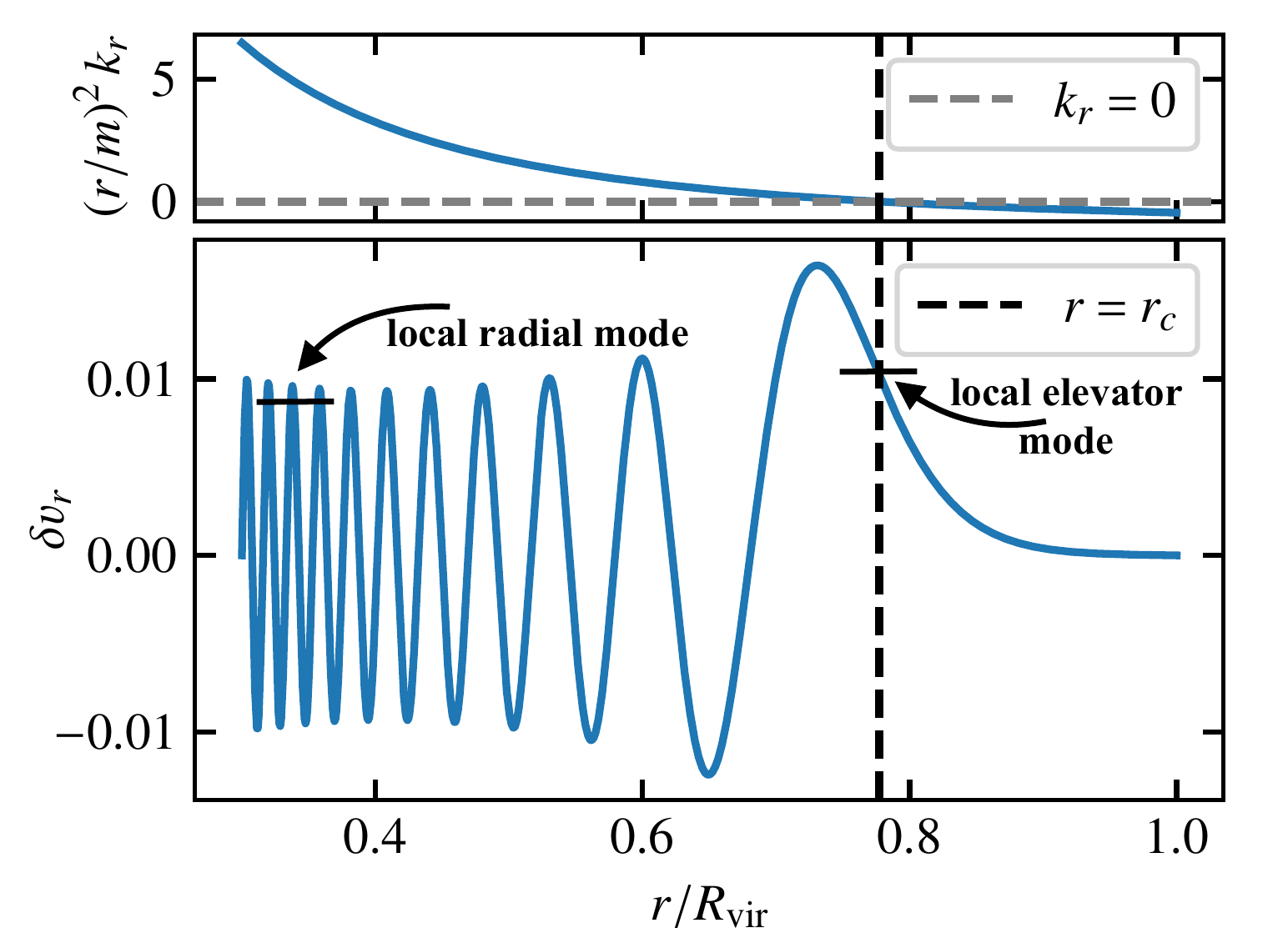}
\caption{
  Illustrative numerical global MTI eigenmode with $m=50$ and $n=20$.
  Top: varying radial wavenumber, $k_r = \sqrt{E - V_\mathrm{ftd}}$, as a function of the radius, $r$.
  Bottom: a high-order global MTI mode.
  The turning point, $\rtp$, of the mode is indicated by black dashed vertical lines.
  It can be found as the single solution to the equation $k_r(r)=0$,
  indicated by the grey dashed horizontal line on the top panel.
}
\label{fig:fideigenmode}
\end{figure}

  We first recall that $\omegaT(r)$ is usually a decreasing function of the radius in the outskirts of galaxy clusters.
  This is also the case in our model of ICM atmosphere, described in \sect{model},
  whose overall MTI frequency profile is displayed in Fig. \ref{fig:profileMTIfrequency}.
  The maximum local growth rate, $\omegaTin$, is thus
  located at the inner boundary, $\rin$, and reached for the local MTI elevator mode there.
  On the contrary, the (smallest) MTI frequency, $\omegaTout$, is reached at the outermost radius, $\rout$.
  To better qualitatively describe the behaviour of the global MTI eigenmodes,
  we rewrote the single governing ODE,
  in the limit of fast thermal diffusion, under Schrödinger form,
\begin{equation}
  \left[
  \frac{\dd^2}{\dd r^2}
- \frac{m^2}{r^2}\left(1 - \frac{\omegaT^2(r)}{\sigma^2}\right) 
  \right]
  \left(r^2\rho_0\delta\varv_r\right)
  = 0.
  \label{eq:ftdschro}
\end{equation}
  From the results obtained in Appendix \ref{app:sl}, we learnt that the growth rate, $\sigma$,
  lies somewhere between $0$ and $\omegaTin$.
  So eigenmodes with $\sigma > \omegaTout$ present
  a turning point at some critical radius, $\rtp$,
  defined as the only solution to $\omegaT(r)=\sigma$ in $r\in\left[\rin,\rout\right]$.
  On the left ($r<\rtp$) of the turning point, the modes oscillate in space with a varying wavenumber, $k_r(r)=\sqrt{E-V_\mathrm{ftd}}$, because their energy,
  $E = (m^2/r^2)(\omegaT^2(r)/\sigma^2)$,
  is higher than the simplified potential,
  $V_\mathrm{ftd} = m^2/r^2$ (where the subscript $\cdot_\mathrm{ftd}$ stands for fast thermal diffusion).
  This radial wavenumber, $k_r(r)$, is displayed on the top panel in Fig. \ref{fig:fideigenmode}.
  On the right ($r>\rtp$) of the turning point though, the modes are evanescent.
  Physically, beyond the turning point,
  the local MTI frequency is no longer large enough
  to support the efficient growth of the global MTI mode,
  which has no other avenue but to decay there.
  For a given azimuthal order, $m$,
  the higher the mode number, $n$, the smaller the growth rate, $\sigma$,
  and hence the further away the turning point
  (and the wider the region of allowed growth, until it reaches the whole domain).
  As illustrated on the bottom panel in Fig. \ref{fig:fideigenmode}, which depicts
  a representative numerically computed MTI eigenmode,
  those modes are necessarily located at smaller radii,
  where the radial wavenumber is positive: $k_r>0$.
  This phenomenological picture is very well in line
  with the radial structure of these global eigenmodes,
  deduced from the Sturm-Liouville theory (details are given in Appendix \ref{app:sl}):
  the MTI modes maximise their power weighted by the local MTI growth rate, \eq{R}.
  Global MTI modes, that are nowhere evanescent, could also develop with $\sigma < \omegaTout$.
  Given their slower growth rate but greater radial extent,
  it is unclear whether such modes are dynamically important.

  This result sheds light on the connection between local modes
  and global modes of the linear MTI:
  at a given radius, the fastest growing local mode,
  known as an elevator mode because it has no radial variation,
  is just a local zoom-in on the single global eigenmode
  that features a turning point at the same radius.
  Around this turning point, the global mode exhibits no further radial variation
  (similarly to the associated local elevator mode) since it becomes evanescent.
  Mathematically speaking, this is due to locally having
  $k_r(\rtp)= 0$.
  On the contrary, global MTI eigenmodes that still exhibit radial structure at a given radius
  grow necessarily slower than the single mode which features a turning point at the same radius.
  Locally, these slower growing global modes correspond to radially varying local modes
  (in contrast to elevator modes).
  This phenomenological discussion is illustrated in Fig. \ref{fig:fideigenmode}.
  This reasoning was in fact inspired by the study of the global linear MRI \citepalias{latter15}.
  In both cases, the fastest growing local modes exhibit no radial variation \citep{balbus01}.
  For the MRI, they are called channel rather than elevator modes.
  Such an analogy could be drawn because both Schrödinger equations,
  \eqeq. (18) in \citetalias{latter15} for the MRI, and \eq{ftdschro} here for the MTI
  (in the limit of fast thermal diffusion), share a very similar structure.
  As a final note, we stress that a similar phenomenology could describe
  the localisation of global (possibly compositional) HBI modes in the ICM \citep{latter12,berlok16b},
  though a formal analysis of the kind led here would be necessary to confirm it.
  In these studies however,
  the Braginskii viscosity further helped to confine the HBI eigenfunctions
  to the lowest altitudes of the ICM atmosphere \citep{latter12}.

\subsubsection{WKBJ solutions}
\label{sec:wkbj}

  To conclude this section on analytical results,
  we provide convenient WKBJ approximations to the growth rates of the solutions of the Schrödinger \eq{ftdschro}
  in the limits of either large azimuthal orders, $m/n\rightarrow\infty$, or large mode numbers, $n/m\rightarrow\infty$.

  Starting with the first limit, $m/n\rightarrow\infty$,
  we found that such global MTI modes are the most unstable and
  feature a turning point at a radius, $\rtp$, very close to the inner boundary, $\rin$.
  Given the decreasing profile of MTI frequency, $\omegaT(r)$,
  the solution is wavy between $\rin$ and $\rtp$, beyond which it becomes evanescent.
  A standard WKBJ estimate within the wavy region, $r\in\left[\rin,\rtp\right]$,
  combined with the quantification conditions issued from the inner and outer BCs,
  used in the limit $m/n\rightarrow\infty$,
  leads to the approximated growth rate,
\begin{equation}
  \sigma_n(m) \approx \omegaTin \left[1 - \left(\frac{3}{2\sqrt{2}} \frac{\left(n + \frac{3}{4}\right)\pi}{m} \left|\frac{\dd\log\omegaT}{\dd\log r}\right|_{r=\rin}\right)^\frac{2}{3}\right].
\label{eq:wkbjcrude}
\end{equation}
  The details of this derivation are given in Appendix \ref{app:crude}.

  Next, we also provide an approximated growth rate for global MTI modes,
  in the limit of large mode numbers, $n/m\rightarrow\infty$. 
  These modes are nowhere evanescent
  in our domain (i.e. $\sigma < \omegaTout$).
  We show in Appendix \ref{app:crude} that the growth rates of these slowly growing MTI eigenmodes 
  can be approximated as
\begin{equation}
  \sigma_n(m) \approx \frac{m}{n\pi} \int_{\rin}^{\rout} \frac{\omegaT}{r} \dd r.
\label{eq:wkbjcrude2}
\end{equation}
  The integral on the right-hand side is easily computed numerically
  and its value depends only on the background structure, not on the particulars of any mode.
  In the next section, these analytical WKBJ approximations, \eq{wkbjcrude}[wkbjcrude2], are compared,
  in the relevant limits, $m/n\rightarrow\infty$ and $n/m\rightarrow\infty$,
  to the growth rates of numerically computed global MTI eigenmodes.

\section{Numerical results}
\label{sec:numerical}
  In this section, we describe our numerical work.
  First, we numerically computed the solutions of the global linear MTI eigenproblem,
  thanks to the Chebyshev pseudo-spectral solver presented in \sect{model}.
  Then, we performed direct numerical simulations of the MTI linear phase with the code IDEFIX,
  to cross-check the validity of the two methods.

\begin{figure*}
\centering
\includegraphics[width=0.85\hsize]{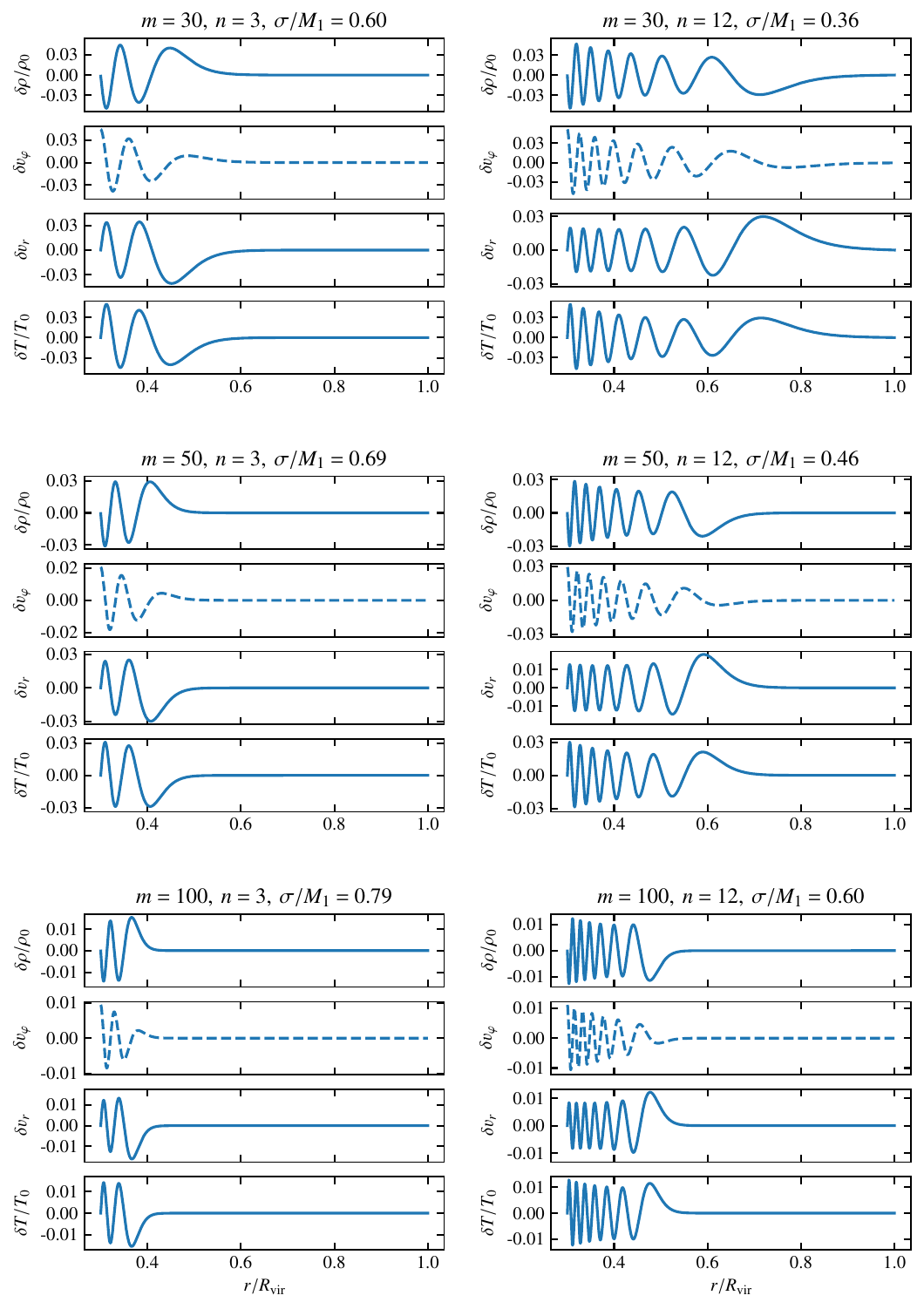}
\caption{
  Numerical solutions to the eigenproblem \eq{drho}[dvphi][dvr][dT][dA].
  From top to bottom: global MTI eigenfunctions for azimuthal order $m=30,50,100$, respectively.
  From left to right: global MTI eigenfunctions for mode number $n=3,12$, respectively.
  From top to bottom in each subfigure: perturbations of density, azimuthal velocity, radial velocity, and temperature
  as a function of the radius.
  The potential vector perturbation is not shown because it simply relates to that of radial velocity through \eq{dA}.
  Real parts of the modes are in plain lines, while imaginary parts are in dashed lines.
  The part whose corresponding line type is absent is zero.
  The modes oscillate in space up to their respective turning points, $\rtp$, and evanescent beyond.
}
\label{fig:numsol}
\end{figure*}

\begin{figure}
\includegraphics[width=\hsize]{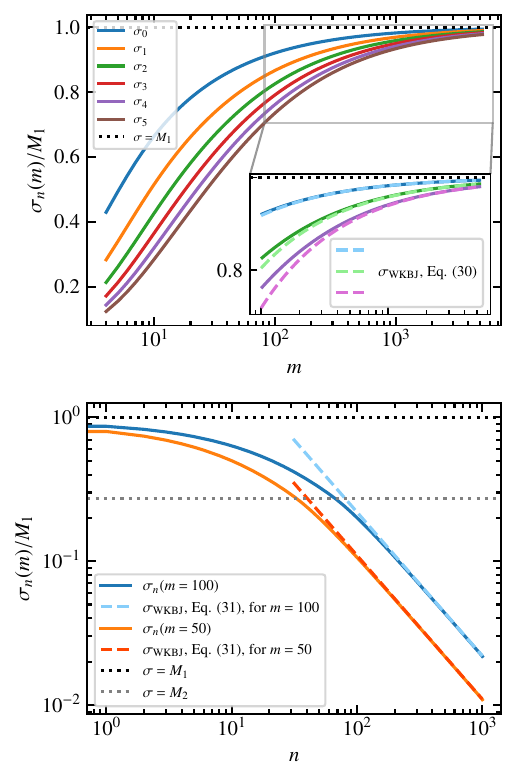}
\caption{
Dispersion relations of the global MTI eigenmodes.
Top: growth rates of the first six MTI modes as a function of the azimuthal order, $m$.
The curve $\sigma=\omegaTin$ is indicated in dotted dark line.
It seems to be a common asymptote to all curves, $\sigma_n(m)$.
Inset: same growth rates, $\sigma_n$, again (one over two for the sake of visibility though),
together with their WKBJ approximations, \eq{wkbjcrude}, in dashed lines.
Bottom: growth rates of the MTI eigenmodes with $m=100$ and $m=50$ 
as a function of the mode number, $n$, in plain lines.
Their WKBJ approximations, \eq{wkbjcrude2}, are shown too, in dashed lines.
The curve $\sigma=\omegaTout$ is indicated in dotted grey line.
}
\label{fig:disp}
\end{figure}

\subsection{Numerical solutions}
\label{sec:numericalcheby}
  We present here some numerical solutions to the system of \eq{drho}[dvphi][dvr][dT][dA]
  for the unknowns $\delta\rho/\rho_0, \delta\varv_\vphi, \delta\varv_r, \delta T/T_0, \delta A/B_0$,
  within a finite domain extending from $\rin=0.3$ to $\rin=1$.
  We stress that the equations solved here do not make use of the sound-proof approximation,
  or of the fast thermal diffusion limit, 
  previously introduced in \sect{analytical} to analytically probe the global MTI eigenmodes.
  A variety of numerically computed MTI eigenmodes are depicted in Fig. \ref{fig:numsol}.
  All of them are normalised so that $Q[r^2\rho_0\delta\varv_r]=1$.
  According to the Sturm-Liouville theory discussed in \sect{analytical},
  the mode number, $n$, at fixed $m$, is also the number of nodes
  displayed by the radial velocity component of the corresponding eigenmode.
  Similarly,
  the different eigenfunctions become evanescent beyond a certain radius,
  given by the mode turning point.
  The location of this turning point depends on the overall mode growth rate:
  most unstable modes are indeed more localised in the innermost regions of the domain,
  and they present a very simple radial structure when they are associated with the fundamental mode number,
  $n=0$ (not shown).
  In contrast, more slowly growing modes extend further out in the atmosphere,
  and they exhibit richer radial structures when they take higher mode numbers, $n>0$.

  Next, we focused on the dispersion relation of the global linear MTI.
  On the top panel in Fig. \ref{fig:disp}, we show the growth rates, $\sigma_n$, of the first six modes
  as a function of the azimuthal order, $m$.
  Appreciable growth of all the illustrated MTI modes occurs for $m\gtrsim30$.
  Interestingly, the growth rates of all modes seem to converge to each other
  in the limit of large azimuthal orders, $m/n\rightarrow\infty$,
  where they reach the asymptote $\sigma\rightarrow\omegaTin$.
  This behaviour is only possible because the present study does not include
  any other diffusive effect, such as isotropic viscosity or magnetic resistivity,
  nor magnetic tension, that would otherwise stabilise the small scales.
  In Fig. \ref{fig:disp} (inset), we plot the WKBJ analytical approximation, \eq{wkbjcrude}, for modes $n=0,\ 2,$ and $4$,
  on top on the numerically obtained growth rates.
  We see that they agree very well with each other, especially in the limit of large azimuthal orders, $m/n\rightarrow\infty$,
  corresponding to modes with turning points, $\rtp$, very close to the inner radius, $\rin$.

  Additionally, we probed the behaviour of the global MTI dispersion relation
  as a function of the mode number, $n$, at fixed azimuthal order, $m$.
  On the bottom panel in Fig. \ref{fig:disp}, we plot the growth rates
  $\sigma_n(m=100)$ and $\sigma_n(m=50)$ for mode numbers up to $n=1000$.
  Such modes exhibit slower growth rates
  ($\sigma_n\rightarrow 0$ as $n\rightarrow \infty$, as expected from the WKBJ approximation)
  but richer radial structure and further extent within the ICM atmosphere,
  since the critical radii of their turning points are far away from the inner boundary, $\rin$.
  In fact, such turning points for global MTI modes with growth rates, $\sigma_n$, lower than $\omegaTout$
  (below the horizontal grey line on the bottom panel in Fig. \ref{fig:disp})
  are outside the numerical solution region, $\left[\rin,\rout\right]$.
  On the same panel, the growth rates, \eq{wkbjcrude2}, approximated with a WKBJ procedure,
  are shown for both $m=100$ and $m=50$.
  The WKBJ approximation, \eq{wkbjcrude2}, perfectly fits
  the numerically obtained growth rates, in the limit of large mode numbers, $n/m\rightarrow\infty$.
  In particular, it captures the right asymptotic behaviour, $\sigma_n \sim n^{-1}$.

\subsection{Direct simulations}
  We present three different numerical simulations of the MTI linear stage,
  performed with the code IDEFIX, described in \sect{model}.
  We first discuss further the numerical implementation of the BCs,
  along with the diagnostics, used in these simulations.
  Then, we assess how well global MTI eigenmodes, and their growth rates, can be reproduced in these simulations,
  by comparing them to the numerical eigenmodes presented in \sect{numericalcheby}.

\subsubsection{Boundary conditions and diagnostics}
  In the light of the common radial structure shared by all global MTI modes seen in Fig. \ref{fig:numsol},
  we describe the BCs imposed in all our simulations.
  For all fields, we took periodic BCs in the azimuthal direction.
  More care was needed to choose the BCs in the radial direction.
  At the inner boundary for instance, the radial velocity, the temperature, and the density perturbations
  of all modes follow a homogeneous Dirichlet condition.
  In contrast, the fractional pressure perturbation is non-zero
  at the inner boundary, while its first derivative must be zero, according to \eq{dvr}.
  Therefore, we paid special attention in preserving these specific traits of the modes
  when designing our inner BCs:
  we either imposed symmetry (i.e. zero gradient) or anti-symmetry of the perturbations around their equilibrium values
  (which are known from the initial HSE condition).
  Regarding the outer boundary,
  we simply let the different fields take their equilibrium values.
  Such outer BCs are only suited for the fastest growing modes,
  which are evanescent beyond a given radius, but not for all MTI eigenmodes.
  This implementation ensures that the BCs disturbed neither the symmetry of the MTI eigenfunctions,
  nor the initial HSE on top of which they developed.

  Moving on to the diagnostics, we systematically probed the radial structure of all the components
  (velocities, pressure, temperature, and density perturbations)
  of the linear modes developing in the simulation domain,
  through a Fourier analysis of their azimuthal dependency,
\begin{equation}
  \delta f_m(r) = \frac{1}{2\pi}\int_{0}^{2\pi} \exp(\id m \vphi)\delta f(r,\vphi)\ \dd\vphi,
\label{eq:modestructure}
\end{equation}
  where $\delta f$ is either $\delta\varv_r,\delta\varv_\vphi,\delta T$ or $\delta\rho$.
  The power contained in each component of the mode was then measured as
\begin{equation}
  \volave{\delta f_m} = \frac{3}{\rout^3 - \rin^3} \int_{\rin}^{\rout} \left|\delta f_m(r)\right| r^2 \dd r.
\label{eq:modepower}
\end{equation}
  The numerical growth rate of each mode was obtained by following the temporal evolution
  of these different powers, over a time window during which this evolution was linear
  (i.e. after efficient growth of the mode started and before it saturated).

  In the next subsection, we present three different simulations,
  all of them initialised with very low amplitude perturbations, of the order of $10^{-10}$.
  The first run used a random white noise on the velocity field and extends from $0$ to $2\pi$ in the azimuthal direction.
  The two other runs were seeded with a single global MTI eigenmode each and extended from $0$ to $\pi$,
  so as to efficiently isolate the growth of a given mode with enough azimuthal resolution.

\subsubsection{Numerical growth rates and MTI eigenfunctions}

\begin{figure}
\includegraphics[width=\hsize]{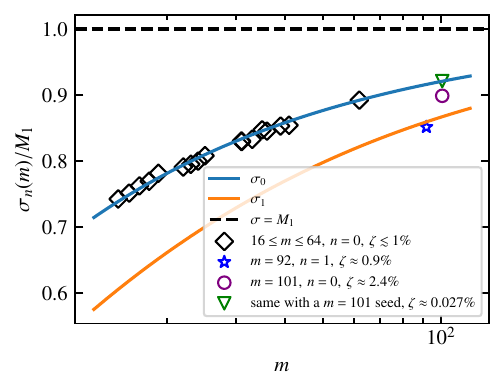}
\caption{
Comparison between the analytical dispersion relation (in plain lines)
and the numerically measured growth rates (markers).
Black diamonds are the numerical growth rates of the most unstable modes with $n=0$,
found in the first simulation seeded with low amplitude random white noise on the velocity field.
They match the theoretical dispersion relation to within $\zeta \lesssim 1\%$.
The blue star depicts the MTI eigenmode with $m=92$ and $n=1$,
which could be successfully isolated in the same simulation.
The purple circle stands for the numerical growth rate of the MTI eigenmode with $m=101$ and $n=0$.
It agrees with the theoretical dispersion relation to within $\zeta\approx 2.0\%$ only.
The green triangle represents the numerical growth rate of the exact same mode,
but in a simulation seeded with a single mode.
The error is very small, $\zeta\approx0.027\%$.
}
\label{fig:numgrowthrates}
\end{figure}

\begin{figure}
\includegraphics[width=\hsize]{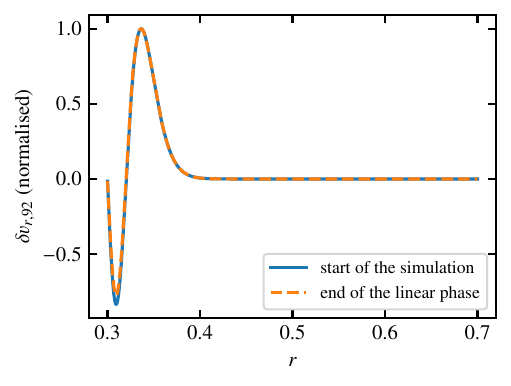}
\caption{
  Radial velocity of the global MTI eigenmode with $m=92$ and $n=1$ in the simulation seeded with a single mode.
  Radial eigenfunctions, $\delta\varv_{r,92}(r)$, are shown at two different times
  and normalised by their maximum values.
  Despite growing by a factor $\sim 10^7$, the eigenmode still lies almost perfectly on top
  of the initial curve, and thus matches very precisely the theoretical eigenfunction.
}
\label{fig:singlemode}
\end{figure}

  We plot a selection of numerical growth rates
  on top of the theoretical dispersion relations for the fastest growing MTI modes in Fig. \ref{fig:numgrowthrates}.
  The chosen numerical growth rates, $\sigma_\mathrm{num}$,
  corresponding to the black diamonds, are measured independently from each other in the first simulation,
  which was initialised with a random white noise.
  They are required to match the theoretical growth rates,
  $\sigma_\mathrm{th}$,
  computed with the Chebyshev pseudo-spectral method, to within $\zeta \lesssim 1\%$
  (the error was computed as $\zeta = \left|\sigma_\mathrm{num} - \sigma_\mathrm{th}\right|/\sigma_\mathrm{th}$).
  Additionally, for each of them, we visually checked that the radial structure of the associated mode
  is easily identifiable and that it matches well the structure of the theoretical eigenmode.
  At the end of this filtering process, 17 fastest growing MTI eigenmodes
  (with $n=0$ and azimuthal orders between $15 \le m \le 62$) are identified in the first simulation.
  Modes with $m \lesssim 14$ can not grow efficiently because the support of their radial eigenfunctions
  extends beyond the outer boundary, $\rout=0.7$, used in the simulations.
  Similarly, we demonstrate below that modes with $m \gtrsim 70$ are not sufficiently well resolved to grow
  at the right rate, at least in the first simulation seeded with random white noise on the velocity field.

  Isolating higher order MTI modes with $n > 0$ in the first simulation is trickier,
  because they grow slower and are thus always supplanted by $n=0$ modes.
  Nevertheless, we managed to identify a persisting MTI eigenmode $n=1,\ m=92$ in this simulation.
  The time evolution of this mode, $\delta\varv_{r,92}(r)$,
  is roughly self-similar (not shown),
  despite growing by a factor $\sim10^{6}$ through the course of the simulation.
  Its growth rate matches that from the theoretical dispersion relation to within $\zeta \approx 0.9\%$,
  so it is added as a blue star in Fig. \ref{fig:numgrowthrates}.
  However, comparing this numerical eigenmode with $m=92$ and $n=1$ against the corresponding theoretical mode
  reveals that its radial structure is not completely resolved at the innermost radii (not shown).
  Therefore, we also produced a simulation with a single seed of the same mode in order to provide an unequivocal comparison basis.
  For this simulation, we decreased the azimuthal extent of our solution domain
  from $2\pi$ to $\pi$, while keeping $4096$ cells in this direction,
  to ensure that the mode is sufficiently well resolved
  and that it is not supplanted by the mode $n=0$ (which can be triggered by discretisation errors) in the course of the simulation.
  In Fig. \ref{fig:singlemode}, the quantity $\delta\varv_{r,92}(r)$ is plotted,
  at two different times during the simulation seeded with a single mode. It is normalised by its current maximum.
  This time, the mode evolution is completely self-similar across time,
  despite increasing by a factor $\sim10^7$.
  Accordingly, the numerically measured growth rate agrees
  with that from the theoretical dispersion relation to within $\zeta \approx 0.002\%$,
  a thousand times better than in the first simulation.

  Finally, we also added by hand the numerical MTI eigenmode with $m=101$ and $n=0$,
  identified in the very first simulation,
  as a purple circle in Fig. \ref{fig:numgrowthrates}.
  This time, while the radial structure of the numerical mode, $\delta\varv_{r,101}(r)$,
  matches quite well that of the theoretical MTI eigenmode (not shown),
  its numerical growth rate is slightly too low with respect to the dispersion relation.
  The discrepancy is of the order of $\zeta\approx2.4\%$.
  We therefore performed a dedicated simulation seeded with a single MTI mode $\left(m=101,n=0\right)$
  and with an azimuthal extent reduced again from $2\pi$ to $\pi$.
  The numerical growth rate measured in this simulation agrees with that from the theoretical dispersion relation
  to within $\zeta\approx0.027\%$, a hundred times better than for the randomly seeded simulation.
  This data point is shown as a green triangle in Fig. \ref{fig:numgrowthrates}.
  We deduced that the azimuthal resolution in the very first run is not high enough to satisfactorily
  model the development of the fastest growing MTI eigenmodes with $m\gtrsim 70$,
  at least when other modes with equivalent azimuthal orders are developing concomitantly.
  Altogether, these results provide a valuable cross-validation of both the linear theory and the numerical codes.

\section{Conclusion}
\label{sec:conclusion}
  In this paper, we have investigated the global linear stability properties
  of dilute stratified plasma subject to anisotropic transport of heat.
  Such a study might be relevant to the intracluster medium (ICM),
  which seemingly meets all the required conditions for the development of the magneto-thermal instability (MTI),
  and which is already known to sustain mild turbulence \citep{hitomi16,hitomi18,xrism25a,xrism25b}.

\subsection{Summary}
  The main objective of this work was to unveil the deeper connections existing
  between local modes and their global counterparts, through a global linear analysis of the MTI in the ICM.
  A similar linear study was previously performed for the magneto-rotational instability (MRI)
  in the context of accretion discs by \citetalias{latter15}.
  We were able to build upon this earlier analysis,
  on account of the very similar linear behaviours shared by the MTI and the MRI \citep{balbus01}.

  Our main findings are summarised below:
\begin{itemize}
\item
  The global linear analyses of both the MRI and the MTI share many similarities.
  The most striking feature is the very similar overall eigenmode structure for these two instabilities.
  In both cases, the fastest growing local modes lack radial structure
  (they are called channel and elevator modes, in the case of the MRI and the MTI, respectively).
  These modes correspond, for both instabilities, to the evanescent parts of global eigenmodes;
  while the slowest growing local modes, which feature non-trivial radial structures,
  are in fact segments of the same global mode, but at smaller radii \citepalias{latter15}.
  This feature may explain why local and global models of ICM atmosphere produce
  similar magneto-thermal turbulence \citepalias{pl22a,pl22b,kempf24}, independently of the scale that they probe.
\item
  From a numerical point of view, we successfully computed numerical solutions to the global MTI eigenproblem
  with our Chebyshev pseudo-spectral solver.
  These solutions were compared with direct numerical simulations of the MTI linear stage
  with the new generation finite-volume code IDEFIX, including a Braginskii module for anisotropic diffusion.
  Therefore, the results presented in this paper provide a valuable benchmark for any
  astrophysical fluid dynamics code implementing the Braginskii heat flux in spherical geometry.
\end{itemize}
  This work is an additional step towards a better understanding
  of how local and global approaches to linear stability theory intertwine.
  It is our hope too that this paper will help to pave the way for a clearer view 
  of the exact role played by global linear eigenmodes,
  both in terms of preferred amplitudes and locations, at turbulent saturation.
  More work is however needed to draw a complete and generic picture of the non-linear saturation
  of such instabilities.
  We briefly comment on this issue in the next subsection.

\subsection{Caveats and future improvements}
\label{sec:B0}
  Although the model developed in this work could satisfactorily address the questions raised in the introduction,
  it could still be improved in many ways.
  First, an obvious refinement of the model would consist in including neglected non-ideal transport effects,
  such as magnetic resistivity or (possibly anisotropic) viscosity,
  or magnetic feedback on the flow through the Lorentz force.
  However, given the specific parameter regime of the ICM in terms of plasma beta,
  and regular and magnetic Prandtl numbers,
  we anticipate that these effects are not critical and would only modestly affect the results presented here.

  Other clear caveats of the current analysis are related to our approximations of the modes being 2D,
  and of the magnetic configuration being purely toroidal, $\vec{B} = B_0\ephi$.
  The main advantage of these assumptions is to leave the model analytically tractable,
  while still capturing two of the main features of the ICM in galaxy clusters,
  namely stratification and anisotropic transport of heat,
  which are critical for the MTI.
  Going 3D and considering the full sphere, $\left(r,\theta,\varphi\right)$,
  by adding the so far neglected polar dependency and component, $\etheta$,
  would be an easy way to implement more complex magnetic geometries
  while preserving the thermal equilibrium of spherically symmetric background profiles.
  Such a development on the full sphere would also enable us to probe
  the polar dependency, out of the equatorial plane, of the global MTI modes in the ICM.
  However, since the Braginskii heat flux is a non-linear operator
  coupling the magnetic field to the temperature field,
  a single mode analysis, on the basis of the (vector) spherical harmonics, as presented by \citet{choudhury16}
  in the case of the thermal instability, is precluded for the 3D linear MTI.
  Such a linear analysis on a basis of spherical harmonics would result in a 2D eigenvalue problem,
  whose solution is intrinsically more demanding than the 1D problem considered in this work.
  Such an extension of our model is outside the scope of the current study.

  Finally, the global linear analysis of the MTI in the ICM carried out in this paper
  is completely oblivious to the specific energetic framework recently developed by \citetalias{kempf24}.
  In that work, \citetalias{kempf24} showed that the notion of available potential energy (APE)
  is anchored at the very heart of the saturation mechanism of magneto-thermal turbulence.
  This phenomenology was inspired by fundamental studies in the context of stratified fluid dynamics
  \citep{middleton20,tailleux24b},
  which suggested that having a net positive conversion rate of APE
  is a necessary condition for the linear development of double-diffusive instabilities,
  such as thermohaline or semi-convection, in the ocean.
  Therefore, it is only natural to wonder whether the same physics,
  namely having a net positive conversion rate of APE,
  could also be responsible for the triggering of linear magnetised buoyancy instabilities in dilute plasma, or not.
  This research direction represents an interesting avenue
  to better constrain the energetics of magnetised buoyancy instabilities in the ICM
  and to further understand how energetics and linear theory interlock.
  Specific work should be devoted to such questions in the future.
  Similarly, the role played by the most unstable MTI modes at saturation remains unclear.
  Future studies carefully examining whether linear theory can be of any help
  for understanding specific features of magneto-thermal turbulence at non-linear saturation,
  especially in terms of the turbulence preferred length scale and intensity \citepalias{pl22a,pl22b,kempf24},
  are therefore needed.

\begin{acknowledgements}
  The authors are thankful to
  Thomas Berlok,
  Joshua Brown,
  Jean-Baptiste Durrive,
  Thomas Jannaud,
  Gordon Ogilvie,
  Prakriti Pal Choudhury,
  François Rincon,
  and Remi Tailleux,
  for many useful discussions which helped to improve the quality of this work.
  It is a pleasure to thank Geoffroy Lesur for sharing
  a version of the balance scheme that he had previously developed in IDEFIX.
  J.M.K acknowledges funding through the grant EUR TESS N°ANR-18-EURE-0018
  in the framework of the Programme des Investissements d'Avenir,
  which funded a three-month stay in Cambridge during his PhD.
  J.M.K is grateful for the hospitality of the host institution, the DAMTP,
  where this work was initiated during summer 2024.
  This work was granted access to the HPC resources of CALMIP under the allocation 2023/2024-P16006,
  J.M.K thanks the associated support team for their work.
  The authors would also like to thank François Rincon for thoroughly reading a draft of the manuscript.
\end{acknowledgements}

%
%

\bibliographystyle{aa}
\bibliography{ref}

\begin{appendix}
\section{Sturm-Liouville theorems}
\label{app:sl}
  In this appendix, we prove the two results stated in \sect{analytical}
  about the growth rates and the radial structures of global MTI eigenmodes.
  This is possible thanks to a specific result of the Sturm-Liouville theory,
  known as the Rayleigh-Ritz rule.
  Before doing so, though, we raise a point of caution that intrinsically limits the scope
  of these results to the most unstable MTI eigenmodes, in the limit of fast thermal diffusion.

  Formally speaking, unveiling an approximation to the pristine \eq{schro}
  under the form of a linear Sturm-Liouville equation, \eq{ftdsl},
  does not automatically ensure that all known results of the Sturm-Liouville theory
  apply \textit{per se} to the global linear theory of the MTI,
  precisely because \eq{ftdsl} is only an approximation.
  In the relevant limit, one might hope however that conclusions relevant to the problem at hand
  can still be found in the results of the Sturm-Liouville theory.
  In the remaining part of this section, we do as if this was the case,
  even though further mathematical work
  would be needed to justify more rigorously the approach taken here.

\subsection{Growth rate}
  First, we derived bounds for the Sturm-Liouville eigenvalue, $\epsilon$.
  We multiplied \eq{ftdsl} by
  $\left(r^2\rho_0\delta\varv_r\right)^*$
  where $\cdot^*$ denotes the complex conjugate,
  and integrated over the solution domain, $\left[\rin,\rout\right]$.
  The first term on the left-hand side was integrated by parts, and the homogeneous Dirichlet BCs 
  were used to cancel the subsequently arising border terms, so as to obtain
\begin{equation}
  \epsilon = \frac{\displaystyle\int_{\rin}^{\rout}\frac{m^2}{r^2}\left|r^2\rho_0\delta\varv_r\right|^2\dd r}{\displaystyle\int_{\rin}^{\rout}\frac{m^2}{r^2}\frac{\omegaT^2(r)}{\omegaTin^2}\left|r^2\rho_0\delta\varv_r\right|^2\dd r} + \frac{\displaystyle\int_{\rin}^{\rout}\left|\ddr\left(r^2\rho_0\delta\varv_r\right)\right|^2\dd r}{\displaystyle\int_{\rin}^{\rout}\frac{m^2}{r^2}\frac{\omegaT^2(r)}{\omegaTin^2}\left|r^2\rho_0\delta\varv_r\right|^2\dd r}.
\label{eq:slbounded}
\end{equation}
  From this expression, we saw that the Sturm-Liouville eigenvalue, $\epsilon$, must be positive,
  provided that the domain is everywhere unstable to the MTI
  (i.e. $\omegaT^2(r) > 0$ for $r\in\left[\rin,\rout\right]$).
  This requirement is, in fact, too strong:
  the eigenvalue, $\epsilon$, is positive
  as long as the common denominator in both fractions is positive.
  In principle, this specific case remains accessible (for some eigenmodes at least),
  even though $\omegaT^2(r) < 0$ in some regions of the domain.
  For such modes (which make the denominators positive),
  another bound can be found since the time scale $\omegaTin$ is
  the maximum value of $\omegaT(r)$ over the domain $r\in\left[\rin,\rout\right]$.
  In this case, the numerator of the first fraction on the right-hand side of \eq{slbounded}
  is automatically greater than the denominator of the same fraction,
  and $\epsilon$ must be greater than unity, since the second fraction is also positive.
  In other words, $\epsilon>1$ means that the growth rate of the global MTI eigenmodes
  cannot be higher than the growth rate of the most unstable local elevator mode in the solution domain
  (i.e. the maximum MTI frequency, $\omegaTin$).
  This feature is satisfying from a physical point of view,
  and is further exploited in \sect{analytical}
  to relate local and global MTI modes.

\subsection{Radial structure}
  Next, we constrained the structure of the solutions to the Sturm-Liouville \eq{ftdsl},
  which are the global MTI eigenmodes in the limit of fast thermal diffusion.
  A specific result of the theory, the Rayleigh-Ritz rule, provides 
  a physical relationship linking the global structure of the different MTI eigenmodes
  to the MTI frequency profile, $\omegaT(r)$.
  This criterion states that the Sturm-Liouville eigenvalue, $\epsilon$,
  is the quotient between the two quadratic forms,
\begin{align}
  Q\left[u\right] &= \int_{\rin}^{\rout} \left(\frac{m^2}{r^2} \left|u\right|^2 + \left|\frac{\dd u}{\dd r}\right|^2\right)\dd r, \\
  R\left[u\right] &= \int_{\rin}^{\rout} \frac{m^2}{r^2}\frac{\omegaT^2(r)}{\omegaTin^2} \left|u\right|^2\dd r,
\end{align}
  which satisfactorily corroborates \eq{slbounded}.
  Additionally, the rule states that the solutions of the Sturm-Liouville \eq{ftdsl} are also
  solutions of the variational problem consisting in minimising the ratio $Q\left[u\right]/R\left[u\right]$,
  amongst all functions, $u$, that satisfy the homogeneous Dirichlet BCs.
  The so obtained eigenmode, $u_0$, has then a minimal associated eigenvalue, $\epsilon_0$,
  and the next smallest eigenvalue is obtained through the minimisation of the same ratio,
  under the additional constraint that
\begin{equation}
  \int_{\rin}^{\rout} \frac{m^2}{r^2} \frac{\omegaT^2(r)}{\omegaTin^2} u^*_0 u\ \dd r = 0,
\label{eq:innerproduct}
\end{equation}
  and so on and so forth.
  We point out that minimising $Q\left[u\right]/R\left[u\right]$
  is equivalent to maximising $R\left[u\right]$ under the constraint that $Q\left[u\right]$ is kept constant.
  In other words, the global MTI eigenmodes,
  in the limit of fast thermal diffusion and at fixed azimuthal order, $m$,
  are exactly the $n$-indexed functions that maximise their power weighted
  by the local maximum MTI growth rate, represented by the quantity
  $R[u_n]$.
  These functions therefore maximise their global growth rate, $\sigma_n$,
  under the constraint of a constant weighted Sobolev norm
  (e.g. $Q[u_n] = 1$).
  The Sturm-Liouville theory indicates that the eigenfunctions, $u_n$,
  form an orthogonal basis under the weighted inner product defined by \eq{innerproduct}.
  Physically, this suggests that the MTI eigenmodes are tapping into the
  background profile of maximum local growth rate at different locations,
  in phase quadrature loosely speaking.

\section{Analytical WKBJ approximations}
\label{app:crude}
  In this appendix, we describe the WKBJ procedure used to deduce
  the analytical approximations, \eq{wkbjcrude}[wkbjcrude2],
  of the growth rates, $\sigma_n(m)$.
  We also detail the derivations when necessary.
  We take $m^{-1}$ as our small WKBJ parameter.

  We first focused on the fastest growing global MTI modes, 
  which feature a turning point very close to the inner boundary, $\rin$.
  The critical radius, $\rtp$, of the turning point is defined as the single solution to the equation $\omegaT(r) = \sigma$.
  Given the decreasing profile of MTI frequency, $\omegaT(r)$,
  the solution is wavy between $\rin$ and $\rtp$, beyond which it becomes evanescent.
  A standard WKBJ estimate is then used within the wavy region, $r\in\left[\rin,\rtp\right]$,
\begin{equation}
  r^2\rho_0\delta\varv_r \propto \cos\left[\int_{r}^{\rtp}\frac{m}{s}\left(\frac{\omegaT^2(s)}{\sigma^2} -  1\right)^{\frac{1}{2}} \dd s+ \frac{\pi}{4}\right].
\label{eq:wkbjansatz}
\end{equation}
  The $\pi/4$ phase shift results from matching the WKBJ ansatz at the turning point,
  when the outer radius, $\rout$, is sent to infinity and the solution asked to vanish there \citep{riley06}.
  The quantification of the growth rate, $\sigma_n$, arises from the requirement
  that the eigenfunctions also cancel at the inner boundary, $\rin$,
\begin{equation}
  \int_{\rin}^{\rtp(\sigma_n)}\frac{m}{r}\left(\frac{\omegaT^2(r)}{\sigma_n^2} - 1\right)^{\frac{1}{2}} \dd r
= \left(n+\frac{3}{4}\right)\pi,
\label{eq:wkbj}
\end{equation}
  where the turning radius, $\rtp$, depends on the eigenvalue, $\sigma_n$, itself.
  Generally, \eq{wkbj} is a non-linear equation of the unknown $\sigma_n$,
  and should therefore be solved numerically.
  However, we show that it is still possible to approximate the growth rate, $\sigma_n$,
  in the limit of large azimuthal orders, $m/n\rightarrow\infty$.
  We consider the change of variable $u = 1 - \omegaTin/\omegaT$.
  Introducing also $\lambda=\omegaTin/\sigma > 1$, we get
\begin{equation}
  \int_{0}^{1-\lambda}\left(\frac{\omegaT^2}{r(u)\omegaTin\omegaT'}\right)\frac{\sqrt{\lambda^2 - \left(1-u\right)^2}}{1-u} \ \dd u
  = m^{-1}\left(n+\frac{3}{4}\right)\pi.
\end{equation}
  We were interested in $\lambda\rightarrow 1$. We wrote $\lambda=1+\varepsilon$, for $0 < \varepsilon\ll 1$.
  The integration range above becomes very small (between $0$ and $-\varepsilon$) and 
  thus we could expand the integrand around $u = 0$,
  that is $r=\rin$ or, equivalently, $r=\rtp$.
  To leading order in $\varepsilon$, noting that in the integral $u \sim \varepsilon$ and that $\omegaT'<0$,
  we obtained
\begin{equation}
  \sqrt{2}\left|\frac{\dd\log\omegaT}{\dd\log r}\right|_{r=\rin}^{-1}\int_{-\varepsilon}^{0}\sqrt{u+\varepsilon} \ \dd u
  = m^{-1}\left(n+\frac{3}{4}\right)\pi.
\end{equation}
  After performing the simple integration,
  we recovered the approximated WKBJ growth rate, \eq{wkbjcrude},
  in the limit of large azimuthal orders, $m/n\rightarrow\infty$.

  Next, we focused on slow radially extended MTI eigenmodes, in the limit of large mode numbers, $n/m\rightarrow\infty$.
  Such modes do not feature turning points in the solution domain,
  so a different WKBJ ansatz is used,
\begin{equation}
  r^2\rho_0\delta\varv_r \propto \cos\left[\int_{\rin}^r\frac{m}{s}\left(\frac{\omegaT^2(s)}{\sigma^2} -  1\right)^{\frac{1}{2}} \dd s + \frac{\pi}{2}\right].
\label{eq:otherwkbjansatz}
\end{equation}
  In this case, the $\pi/2$ phase shift results from requiring that the solution vanishes at the inner boundary,
  $r=\rin$,
  and the quantification of the growth rate arises from the outer BC,
\begin{equation}
  \int_{\rin}^{\rout}\frac{m}{r}\left(\frac{\omegaT^2(r)}{\sigma_n^2} - 1\right)^{\frac{1}{2}} \dd r
= n\pi.
\end{equation}
  However, in the limit $n/m\rightarrow\infty$, $\omegaT/\sigma_n\rightarrow\infty$ everywhere in the domain too.
  In this case, the square root appearing within the integrand on the left-hand side of the latter equation is approximated as
  $\omegaT/\sigma_n$ and \eq{wkbjcrude2} directly follows.
  We note that the WKBJ ansatz, \eq{otherwkbjansatz},
  still requires $m^{-1}$ to be a small parameter.
  Our result is then only valid for $n\gg m \gg 1$.
  This is the reason why we used the azimuthal orders $m=50$ and $m=100$,
  together with mode numbers as large as $n=10^3$,
  on the bottom panel in Fig. \ref{fig:disp}.

  Finally, for completeness, we provide a last analytical formula,
  which is valid for MTI modes in the limit $n/m\rightarrow\infty$,
  but when the outer boundary of the domain is sent far away and the MTI frequency required to vanish there:
  $\omegaT\rightarrow 0$ when $r\rightarrow\infty$.
  Modes in such an ICM atmosphere systematically feature a turning point.
  Even though their critical radii, $\rtp$, are located very far away from the inner boundary, $\rin$,
  the same WKBJ ansatz as in the limit $m/n\rightarrow\infty$, \eq{wkbjansatz},
  and the same quantification condition, \eq{wkbj}, can be used.
  In this case however, we were instead interested in the limit $\lambda^{-1} = \sigma/\omegaTin\rightarrow 0$.
  We then used the change of variable $u=\sigma/\omegaT$,
  so that \eq{wkbj} transforms to
\begin{equation}
  \int_{\lambda^{-1}}^{1}\left|\frac{\dd\log\omegaT}{\dd\log r}\right|^{-1}\frac{\sqrt{1-u^2}}{u^2} \ \dd u
  = m^{-1}\left(n+\frac{3}{4}\right)\pi.
\end{equation}
  The integral on the left-hand side is controlled by the nearly diverging $u^{-2}$ factor,
  as the lower integration bound is $\lambda^{-1}\rightarrow 0$.
  The integral is then dominated by the contribution of a small innermost region around $u=\lambda^{-1}$,
  that is near $r=\rin$.
  Assuming that, within this region, the factor $\left|\dd\log\omegaT/\dd\log r\right|$ does not vary significantly 
  and that the second factor is controlled by $u^{-2}$,
  we obtained the approximated WKBJ growth rate,
\begin{equation}
  \sigma_n(m) \approx \omegaTin \frac{m}{\left(n+3/4\right)\pi} \left|\frac{\dd\log\omegaT}{\dd\log r}\right|_{r=\rin}^{-1}.
\label{eq:verylast}
\end{equation}
  Both \eq{wkbjcrude2} and \eq{verylast} share a similar scaling, $\propto n^{-1}$,
  although the physical prefactors in both equations differ.

\end{appendix}
\end{document}